\documentclass[preprintnumbers,showpacs,nofootinbib,superscriptaddress,leqn,twocolumn]{revtex4}
\usepackage{dcolumn}
\usepackage{amsmath,amssymb,setspace,graphicx,subfigure,epsfig}
\usepackage{amsfonts}
\usepackage{latexsym}
\usepackage{epsfig}
\usepackage{color}

\newcommand{\dis}[1]{\begin{equation}\begin{split}#1\end{split}\end{equation}}

\newcommand{\be}{\begin{equation}}
\newcommand{\ee}{\end{equation}}

\def    \bea           	{\begin{eqnarray}}
\def    \eea           	{\end{eqnarray}}

\newcommand{\tev}{\,\textrm{TeV}}
\newcommand{\gev}{\,\textrm{GeV}}

\newcommand{\fermi}{{\textrm{Fermi-LAT} }}
\newcommand{\pamela}{{\textrm{PAMELA} }}
\newcommand{\electron}{{e^-  }}
\newcommand{\positron}{{e^+  }}
\newcommand{\fraction}{{\frac{\Phi_{\positron}}{\Phi_{\electron} +\Phi_{\positron}}}}
\newcommand{\sigmav}{\langle\sigma v\rangle}

\begin{document}

\title{Implications of an astrophysical interpretation of PAMELA and Fermi-LAT data for future searches of a positron signal from dark matter annihilations}
\author{Ki-Young Choi}
\email{kiyoung.choi@pusan.ac.kr}
\affiliation{Departamento de Fisica Teorica and Instituto de Fisica Teorica UAM-CSIC\\
Universidad Autonoma de Madrid, Cantoblanco, E-28049 Madrid, Spain}
\affiliation{Department of Physics, Pusan National University, Busan 609-735, Korea}
\author{Carlos E. Yaguna}
\email{carlos.yaguna@uam.es}
\affiliation{Departamento de Fisica Teorica and Instituto de Fisica Teorica UAM-CSIC\\
Universidad Autonoma de Madrid, Cantoblanco, E-28049 Madrid, Spain}

\begin{abstract}
The recent data from PAMELA and Fermi-LAT can be interpreted as evidence of new astrophysical sources of high energy  positrons. In that case, such astrophysical positrons constitute an additional background against the positrons from dark matter annihilation. In this paper, we study the effect of that  background on the prospects for the detection of a positron dark matter signal in future experiments. In particular, we determine the new regions in the (mass, $\sigmav$) plane that are detectable by the AMS-02 experiment for several dark matter scenarios and different propagation models. We find that, due to the increased background, these regions feature annihilation rates that are up to a factor or three larger  than those  obtained for the conventional background. That is, an astrophysical interpretation of the present data by PAMELA and Fermi-LAT implies that the detection of positrons from dark matter annihilation is slightly more challenging than previously believed.
\end{abstract}

\maketitle

\section{Introduction}
If dark matter consists of WIMPs --weakly interacting massive particles-- they can annihilate with each other and be \emph{indirectly} detected by searching for their annihilation products, mainly gamma rays, neutrinos, and antimatter. This indirect detection technique is in fact one of the promising avenues toward the determination of the dark matter nature. Such a complex task will certainly require input from accelerator searches, direct detection experiments, as well as the indirect detection of dark matter in multiple channels.  Several experiments, including PAMELA \cite{pamela} and Fermi \cite{fermi}, are already looking for such signals in cosmic rays, and planned experiments,  such as AMS-02 \cite{ams2}, will measure the cosmic ray spectrum with even higher precision, possibly revealing a dark matter signature. Thanks to these experimental efforts the  detection of dark matter and its identification may soon become a reality.

Recently, the PAMELA collaboration reported~\cite{Adriani:2008zr} an excess (with respect to the conventional background model) in the  positron fraction for the energy range $1.5-100\gev$. Later on, the Fermi-LAT collaboration reported~\cite{Abdo:2009zk} the measurement of the flux of electrons plus positrons up to $1$ \tev. These two measurements strongly indicate  the existence of additional sources of high energy cosmic ray positrons~\cite{Grasso:2009ma}. Among the possible sources, dark matter annihilations \cite{Meade:2009iu} (or decays) as well as astrophysical sources such as pulsars\cite{Atoian:1995ux,Grimani:2004qm,Buesching:2008hr,Yuksel:2008rf,Hooper:2008kg,Profumo:2008ms} or supernova remnants~\cite{Blasi:2009hv,Blasi:2009bd,Fujita:2009wk} have been considered in the literature. 

Though viable, an interpretation of PAMELA and Fermi-LAT data in terms of  dark matter annihilations is rather disfavoured~\cite{Grasso:2009ma}. In fact, the annihilation rate for a thermal relic  is about three orders of magnitude smaller than the one needed to explain the data. Hence, one has to resort to non-thermal candidates or to low-velocity enhancement mechanisms, both tightly constrained by experimental data \cite{Bertone:2008xr,Cirelli:2009vg,Profumo:2009uf,Galli:2009zc,Slatyer:2009yq}.  Moreover, because both leptons and hadrons are typically generated in dark matter annihilations, they tend to overproduce  gamma rays and antiprotons. As a result, significant regions of the parameter space that is compatible with the positron excess are already excluded by present constraints~\cite{Cirelli:2008pk,Pato:2009fn}. It was also  suggested \cite{Nardi:2008ix,Meade:2009iu} that dark matter decays, rather than annihilations, may explain the positron data. Several models of this kind have been recently studied --see e.g. \cite{Mardon:2009gw}-- and they are typically less constrained than annihilation models. In any case, we are interested in alternative interpretations of the data so we will not consider the dark matter possibility any further.

The additional source of positrons required to explain the PAMELA and Fermi-LAT data may also be of an astrophysical nature, such as pulsars~\cite{Atoian:1995ux,Grimani:2004qm,Buesching:2008hr,Yuksel:2008rf,Hooper:2008kg,Profumo:2008ms} or supernova remnants~\cite{Blasi:2009hv,Blasi:2009bd,Fujita:2009wk}. Indeed, it has been shown that, under reasonable assumptions, they can account for the spectral features observed by Fermi-LAT and for the rising positron fraction measured by PAMELA. In this paper we suppose that the cosmic ray data from PAMELA and Fermi-LAT is explained by new astrophysical sources of high energy positrons.

Such astrophysical positrons constitute,  therefore, a new background against the positrons from dark matter annihilation. This new  background  must be taken into account in determining the sensitivity of future experiments to a positron dark matter signal. That is precisely what we do in this paper. We reassess the sensitivity of the AMS-02 experiment to a positron signal from dark matter annihilations in light of the new positron background implied by an astrophysical explanation of the PAMELA and Fermi-LAT data.

In the next section we introduce the four dark matter scenarios used as benchmark  throughout this paper, and the assumptions that enter into the computation of the positron signal from dark matter annihilation are explained. Then, in section \ref{sec:back}, we derive the positron background implied by an astrophysical interpretation of PAMELA and Fermi-LAT data. Finally, in section \ref{sec:ams}, we obtain our main results. By combining the previously found signal and background, we study the sensitivity of AMS-02 to a positron signal from dark matter annihilations and find the new detectable regions in the plane ($m,\sigmav$) for different dark matter scenarios and propagation models. To isolate the effect of the new background, we also determine how much different these regions are from those obtained for the conventional background model. 

\section{The positron flux from dark matter annihilations}
Positrons can be produced in dark matter annihilations through a variety of different channels. Annihilations into gauge bosons, for instance, will yield positrons as the gauge bosons decay. Decays into leptons will readily produce positrons and those into hadrons  also produce positrons via charged pions. Similarly, annihilations into $b$ quarks will yield positrons through the quark decay into charged leptons. Each of these annihilation modes  gives rise to a specific positron spectrum. To maintain our discussion as general as possible, instead of restricting ourselves to a specific dark matter model or to a given annihilation channel, we will consider four different benchmark models that span a wide variety of possibilities considered in the literature. They are
\begin{enumerate}
\item Models with annihilation into $b\bar b$.  Supersymmetric models with bino-like neutralinos, such as the CMSSM, are the prototype scenario for this final state.
\item Models with annihilation into $W^+W^-$. They include models with a singlet~\cite{Yaguna:2008hd} or a doublet scalar \cite{LopezHonorez:2006gr}. Another example are supersymmetric models with wino-like neutralinos. 
\item Models with annihilation into $e^+e^-$. Models of these type have been recently proposed. They arise, for instance, in scenarios where the dark matter sector is secluded \cite{Pospelov:2007mp}. 
\item A typical model of Kaluza-Klein dark matter. For this model we take the following annihilation branching ratios \cite{Hooper:2007qk}: $20\%$ into each charged lepton,  $11\%$ into each up-type quark, $1.2\%$ into each neutrino, $2.3\%$ into Higgs bosons, and $0.7\%$ into down-type quarks. Consistency with electroweak precision data requires the mass of the dark matter candidate to be larger than $300\gev$.
\end{enumerate}
To obtain the positron spectrum for each of these possibilities we relied on PYTHIA~\cite{Sjostrand:2000wi} as implemented in the DarkSUSY package~\cite{Gondolo:2004sc}.  
\begin{center}
\begin{table}
\centering
\begin{tabular}{|c|ccc|}
\hline 
&$L$ (kpc)&$K_0$(kpc$^2$/Myr)&$\alpha$\\
\hline 
MIN & $1$ & $0.00595$ & $0.55$ \\ 
MED & $4$ & $0.0112$ & $0.70$ \\
MAX & $15$ & $0.0765$ & $0.46$ \\
\hline 
\end{tabular}
\caption{{\footnotesize Values of propagation parameters
widely used in the literature and roughly providing minimal and maximal positron fluxes,
or constitute the best fit to the B/C data.}}
\label{tab:par}
\end{table}
\end{center}
The positrons produced from dark matter annihilation in the Galactic halo lose their energy via inverse Compton and synchrotron processes as they propagate in interstellar space. Such effects can be taken into account by solving the diffusion-loss equation:
\begin{equation}
\frac{\partial}{\partial t}\frac{dn}{dE}=\vec{\nabla}\cdot\left[K(E,\vec{x})\vec{\nabla}\frac{dn}{dE}\right]+\frac{\partial}{\partial E}\left[b(E,\vec{x})\frac{dn}{dE}\right]+Q(E,\vec{x})\,
\end{equation}
where $K$ is the diffusion constant, $b$ is the energy loss rate, and $Q$ is the source term. It is assumed that, within the \emph{diffusion zone},  $K$ is constant in space and  varies only with energy. This energy dependence can be modeled as
\begin{equation}
K(E)=K_0\left(\frac{E}{\mathrm{GeV}}\right)^\alpha \mathrm{cm}^2 \mathrm{s}^{-1}\,,
\end{equation}
where $K_0$ and $\alpha$ are propagation parameters. The energy loss rate is given by
\begin{equation}
b(E)=10^{-16} \left(\frac{E}{\mathrm{GeV}}\right)^2 \mathrm{s}^{-1}\,.
\end{equation}
The source term, $Q$, takes into account the properties of the dark matter particle: its mass, annihilation cross section, dominant annihilation modes, and distribution in the Galaxy. In relation to this last issue, we assume a NFW profile with a local dark matter density of $0.3\gev\mathrm{cm}^{-3}$. Regarding the boundary conditions, we take the diffusion zone to be a slab of thickness $2L$,  and assume that the positron density is zero at $z=\pm L$ -- free escape boundary conditions.  The values of $L$, $K_0$, and $\alpha$ will depend on the propagation model. In the following we will consider the well-known MIN, MED, and MAX models \cite{Delahaye:2007fr}, which assign values to these parameters according to Table \ref{tab:par}. With these elements we can proceed to compute the positron flux from dark matter annihilations.

\begin{figure}[t]
\includegraphics[scale=0.3]{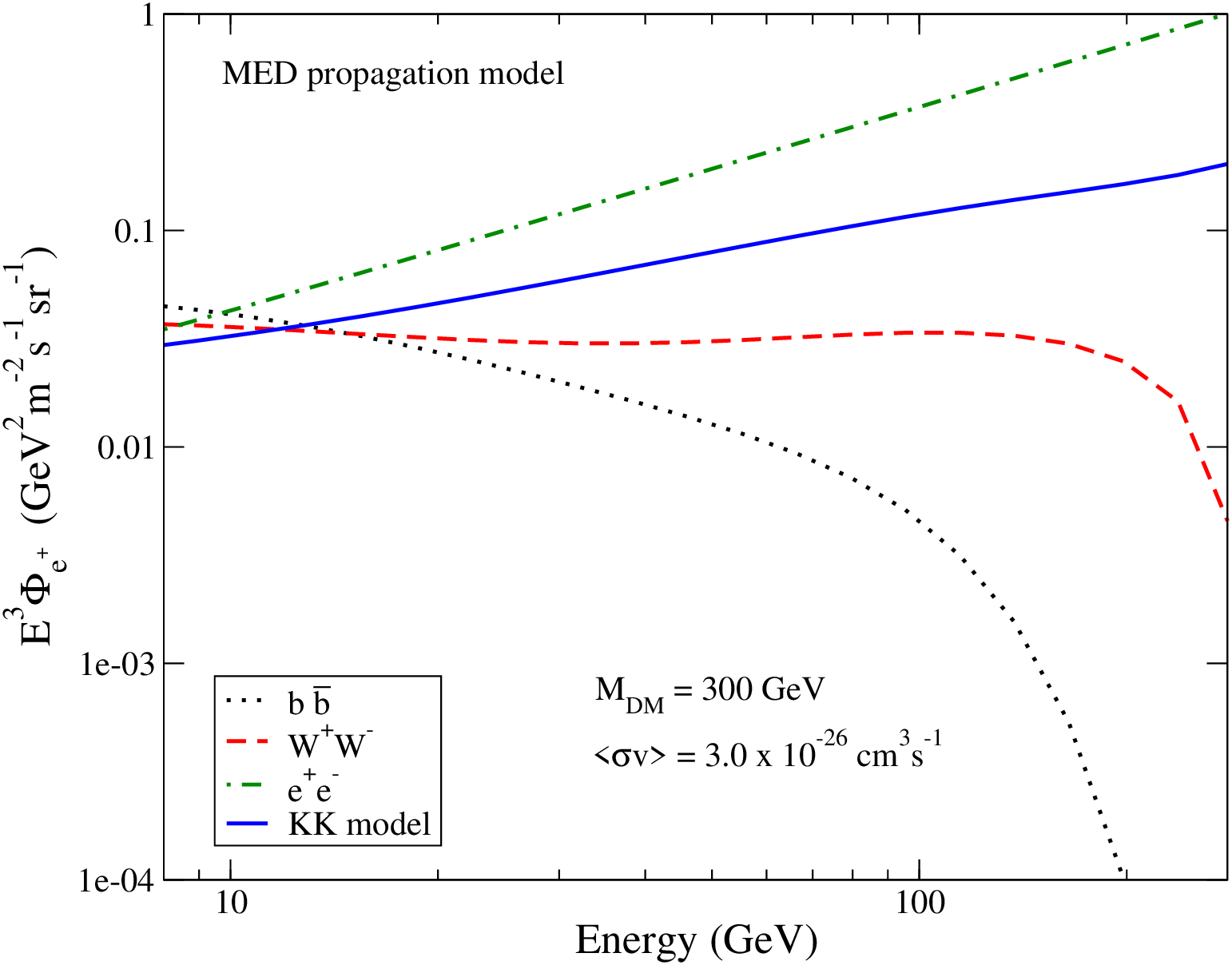} 
\caption{The positron spectrum from dark matter annihilations after including propagation effects (MED model). Each line corresponds to a different model: annihilation into $b\bar{b}$, $W^+W^-$ or $e^+e^-$ and a typical Kaluza-Klein model. In all cases we set $m_{DM}=300\gev$ and $\langle\sigma v\rangle=3.0\times 10^{-26}.$\label{fig:spec}}
\end{figure}

Figure \ref{fig:spec} displays the positron flux as a function of the energy for the  different dark matter scenarios we consider and for the MED propagation model. In that figure the dark matter particle mass and $\sigmav$ were set respectively at $300\gev$ and $3.0\times 10^{-26}\mathrm{cm^3s^{-1}}$. Notice that, at high energies, the largest positron flux  is obtained for the model with direct annihilation into $\positron\electron$, followed in decreasing order by the KK model, the model with annihilation into $W^+W^-$, and the model with annihilation into $b\bar b$, which gives the smallest flux in that energy range. As we will see in section \ref{sec:ams}, these spectral differences have important implications for the future detection of a positron signal.

\section{The new positron background}
\label{sec:back}
To determine the sensitivity of a given experiment to positrons from dark matter annihilation, we need to know not only the signal --the positron flux that we computed in the previous section-- but also the expected background --the positron flux from other sources. Since experiments such as PAMELA and AMS-02 are going to measure the positron flux up to energies of about $300\gev$, we also need to know the background up to those energies. In this section we derive the new positron background implied by an astrophysical interpretation of the PAMELA and Fermi-LAT data.

In the conventional background model, positrons in our Galaxy are produced when cosmic-ray nuclei interact with the interstellar medium. The resulting secondary positron flux can be parametrized as~\cite{Strong:2004de,Baltz:1998xv}
\be
\Phi_\positron^{conv}=\frac{4.5E^{0.7}}{1+650E^{2.3}+1500E^{4.2}}\gev^{-1}\mathrm{cm}^{-2}\mathrm{s}^{-1}\mathrm{sr}^{-1}\,
\ee 
where $E$ is given in GeV. This conventional background, however, is not compatible with the recent measurements by PAMELA and Fermi-LAT. In fact, even after taking into account the uncertainties in the secondary positron flux from cosmic ray propagation, the data reveals a clear positron excess over the expected background \cite{Grasso:2009ma,Balazs:2009wm,Delahaye:2008ua}. Thus, a new  source of high energy positrons  is necessary to explain the data.

In this paper, we  assume that this new source of positrons is \emph{not} related to dark matter  but instead supposed it to be  of astrophysical nature.  Within that framework, the search for positrons from dark matter annihilations remains a challenge for future experiments, and the \emph{astrophysical} positrons revealed by the data contribute to the background in those searches.  

The recent data from PAMELA and Fermi-LAT go a long way toward the determination of the background positron flux. 
The PAMELA collaboration reported~\cite{Adriani:2008zr} the measurement of the positron fraction, $\fraction$, in the energy range between $1.5\gev$ and $100\gev$. The Fermi-LAT collaboration, later on, reported \cite{Abdo:2009zk} the measurement of the total flux of electrons plus positrons, $\Phi_e^-+\Phi_e^+$, in  the $20\gev$ to $1\tev$ energy range. By combining these two measurements, the absolute flux of positrons  can be determined as 
\be
\Phi_\positron=\left(\frac{\Phi_\positron}{\Phi_\electron+\Phi_\positron}\right)_\pamela\times\left(\Phi_\electron+\Phi_\positron\right)_\fermi
\ee
for energies below $100\gev$. Let us emphasize that this procedure became possible only recently, after the release of the electron plus positron spectrum by the Fermi-LAT collaboration. The positron flux up to $100\gev$ can, therefore, be directly extracted from the available data~\cite{Balazs:2009wm}.

\begin{figure}[t]
\includegraphics[scale=0.35]{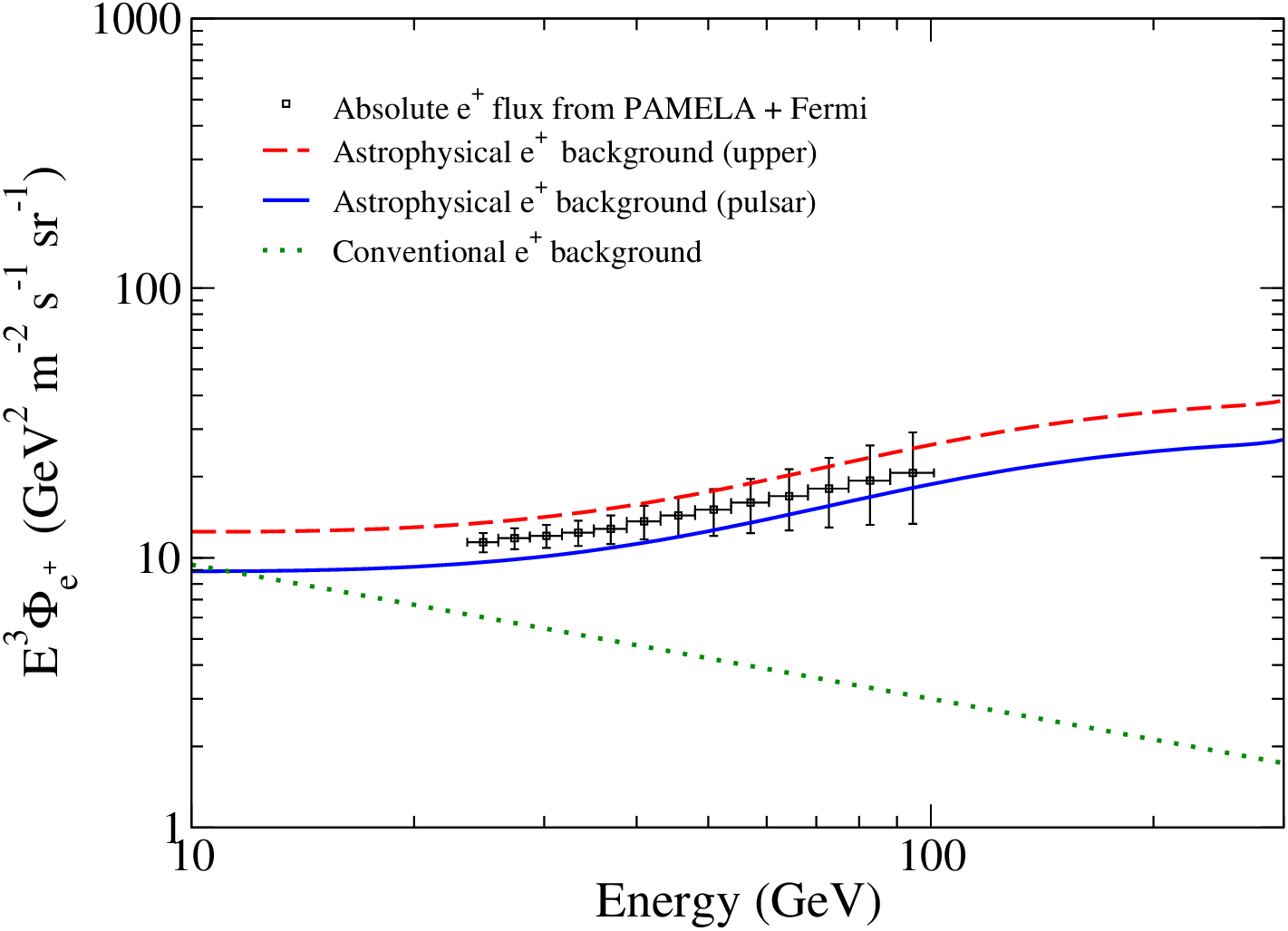} 
\caption{Comparison between the conventional positron background (green dotted line), the absolute positron flux derived from Pamela and Fermi-LAT (black squares),  and the backgrounds suggested by an astrophysical interpretation of PAMELA and Fermi-LAT (blue solid line and red dashed line). See text for details.
\label{fig:back}}
\end{figure}

Beyond $100\gev$ we cannot rely on experimental data to obtain the positron flux, so we will instead resort to a specific astrophysical model: the pulsar interpretation of PAMELA and Fermi-LAT~\cite{Grasso:2009ma}. Such interpretation assumes that in the energy range between $100\gev$ and $1\tev$ the electron and positron flux reaching the Earth is the result  of two different contributions. The first one is an almost homogeneous and isotropic Galactic cosmic ray component produced by supernova remnants. The second one is the local contribution of a few pulsars. This pulsar contribution is expected to be more significant at higher energies. 

It was found in \cite{Grasso:2009ma} that among the candidate sources from the ATNF radio pulsar catalogue, only the Monogem (PSR B0656+14) and the Geminga (PSR J0633+1746) pulsars give a significant contribution to the high energy electron and positron flux. Moreover, a good fit to both the PAMELA and Fermi-LAT data could be obtained with $E_{cut}=1100\gev$, $\eta_{e^\pm}=40\%$, and $\Delta t = 6\times 10^4$yr. We will consider those  parameters as defining the default pulsar model and use them to obtain, from figures 4 and 5 in \cite{Grasso:2009ma}, the positron flux up to  $300\gev$. In figure \ref{fig:back} this positron flux (blue solid line) is compared with the conventional positron background (green dotted line) and also with the absolute positron flux derived from Pamela and Fermi-LAT data in \cite{Balazs:2009wm} (black squares). As seen in the figure, the uncertainty in present data allows the positron flux to be somewhat larger than the prediction of the default pulsar model. To take that possibility into account we consider an additional astrophysical background, denoted in the figure as \emph{upper} (red dashed line). That background has the same shape as the pulsar one but a slighty larger normalization --by a factor $1.4$. In the following, we  refer to these two backgrounds as the astrophysical backgrounds and assume that the true background is somewhere between them. Notice that besides being compatible with the absolute positron flux, the astrophysical backgrounds are both clearly distinguishable from the conventional one.

It is important to stress that our results do not strongly depend on the specific pulsar model adopted and that similar conclusions are obtained even if alternative astrophysical explanations are considered. In fact, up to $100$ GeV the absolute positron flux is essentially known from data, so all models should give roughly the same background in that region. And it is precisely in this low energy region where the positron flux is larger for the dark matter models and masses we consider. Besides, between $100$ GeV and $300$ GeV, the positron flux, even if not known, is not expected to have any particular features, for the $\electron+\positron$ spectrum measured by Fermi-LAT is rather flat. So, in this \emph{high} energy the positron flux is not expected to deviate much from the two backgrounds we study. Moreover, we have explicitly checked that alternative astrophysical interpretations such as that put forward in \cite{Blasi:2009hv} indeed give rise to a very similar positron flux in that energy range. At even higher energies ($E\gtrsim 300\gev$), the predicted flux could well be very different, but that would not affect the detection of a dark matter positron signal in AMS-02.

The  difference between the conventional background and the astrophysical ones depends on the energy and, as seen in figure \ref{fig:back}, amounts to more than one order of magnitude at $300\gev$. This energy dependence is crucial for our analysis. Had the difference between the conventional and the astrophysical backgrounds been an energy independent factor $f$, most of our analysis would have been unnecessary. In that case, the positron dark matter signal required for detection would simply have been $\sqrt{f}$ larger than for the conventional background, independently of the dark matter model. Alas, that is  clearly not the case; the astrophysical backgrounds are not only larger than the conventional one, they also have a different energy dependence. Thus, it is not possible to  predict how much larger the signal will have to be in order to ensure detectability, and such a number is expected to depend also on the dark matter model. In the next section we will quantitatively study these issues.

\section{Sensitivity of AMS-02}
\label{sec:ams}
For definiteness, we will focus our sensitivity analysis on the AMS-02 experiment~\cite{ams2}. AMS-02 will take data for about $3$ years and is expected to measure the positron spectrum up to about  $300\gev$. Its larger  geometric acceptance -- $420\mathrm{cm}^2 \mathrm{sr}$ compared to PAMELA's $20.5\mathrm{cm}^2\mathrm{sr}$ -- make it an ideal instrument to search for and detect the positron signal from dark matter annihilations. 
\begin{figure}[t]
\includegraphics[scale=0.3]{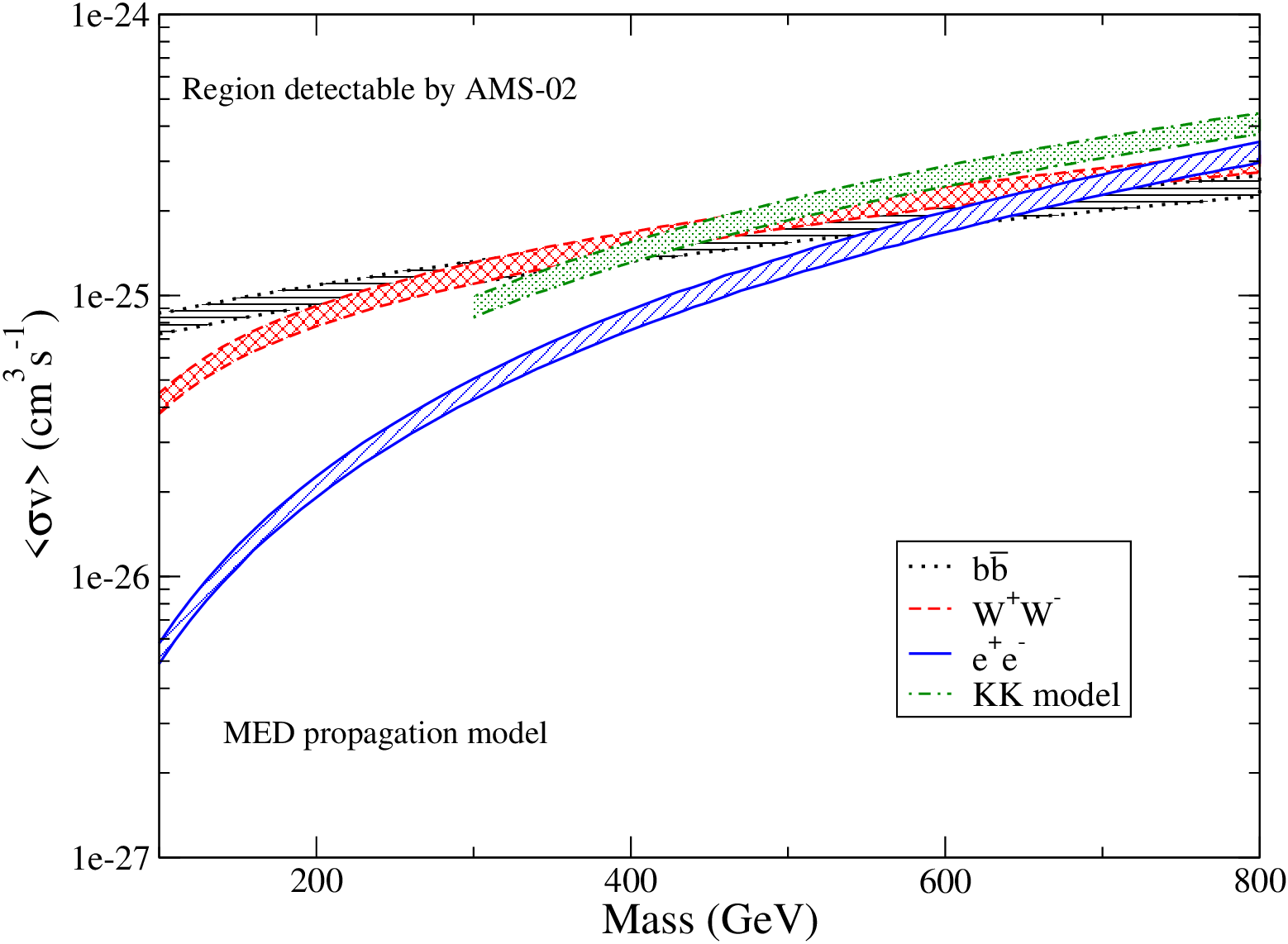} 
\caption{Regions in the plane $(mass, \langle\sigma v\rangle)$ that will be detectable by AMS-02 with three years of data for different dark matter scenarios and the MED propagation model. The area above the band is detectable for both astrophysical backgrounds. \label{fig:med}}
\end{figure}

To claim the observation of a positron signature from dark matter annihilation in a given experiment, the spectral data must be statistically distinguishable from the predicted background spectrum. A $\chi^2$ test is one statistical tool that allows us to assess that hypothesis. The $\chi^2$ of a set of data over the expected background is defined as
\dis{
\chi^2 =\sum_i\frac{(N_i^{obs}-N_i^{BG})^2}{N_i^{obs}},\label{chi2}
}
where the sum is over energy bins, $N_i^{obs}$ is the number of events observed in the $i$ energy bin and  $N_i^{BG}$ is the number of events in the $i$ energy bin predicted from the background contribution. In writing equation (\ref{chi2}), Gaussian errors have been assumed. For our analysis we use  $20$ energy bins equally spaced in logarithmic scale between $10\gev$ and $300\gev$. To assess the sensitivity of AMS-02 we calculate the annihilation rate needed to distinguish, with three years of data,  the signal from the  background at $95\%$ confidence level -- that is, we require that  $\chi^2\gtrsim 31$. Let us emphasize that sensitivity analysis of this kind have been presented in the literature for several models, see for instance ~\cite{Baltz:1998xv,Hooper:2004bq}. They were done, however, before the release of PAMELA and Fermi-LAT data and relied on the conventional positron background. The novelty in our analysis is the use of the new positron  background implied by an astrophysical explanation of PAMELA and Fermi-LAT. 
\begin{figure}[t]
\includegraphics[scale=0.3]{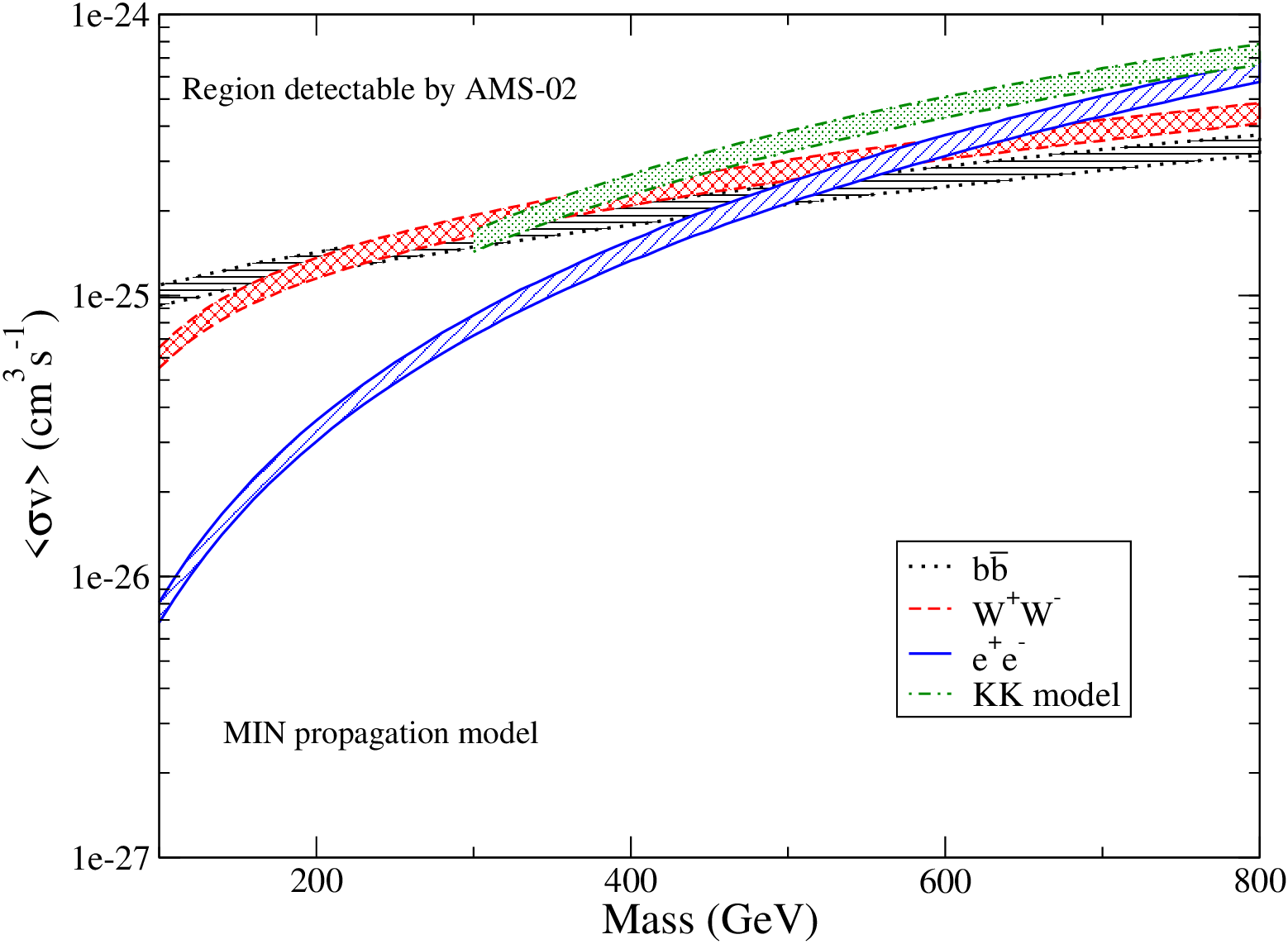} 
\caption{Regions in the plane $(mass, \langle\sigma v\rangle)$ that will be detectable by AMS-02 with three years of data for different dark matter scenarios and the MIN propagation model. The area above the band is detectable for both astrophysical backgrounds. \label{fig:min}}
\end{figure}

Figure \ref{fig:med} shows, in the plane ($mass, \sigmav$), the regions that are detectable by the AMS-02 experiment for the four different dark matter scenarios we consider and for the MED propagation model. For each dark matter model, we compute, at a given mass, the $\sigmav$ required to observe positrons for each of the two astrophysical backgrounds.  A narrow band in the plane ($mass, \sigmav$) is thus obtained for each model. The lower limit of such band corresponds to the results for the astrophysical background from the default pulsar model whereas the upper limit corresponds to the results for the \emph{upper} astrophysical background.  The area above the band can be considered, henceforth, as detectable for both astrophysical backgrounds. From the figure we see that the model with direct annihilation into $e^+e^-$ offers the best perspectives, with required annihilation rates well below the typical one ($\sigmav\sim 3\times 10^{-26}$cm$^{-3}$s$^{-1}$) for low masses, $m\lesssim 300\gev$. The models with annihilation into $W^+W^-$ and $b\bar b$ also look promising. For masses below about $300\gev$, they are detectable even for annihilation rates smaller than $10^{-25}$cm$^3$s$^{-1}$. So, for typical rates only moderate boost factors,  which may be compatible with the standard $\Lambda$CDM scenario  of structure formation~\cite{Lavalle:1900wn}, are needed to enhance the signal to a detectable level.
 A similar conclusion applies to the  KK model in the low mass range, $m\lesssim 350\gev$. Thus, even if not guaranteed, AMS-02 has good chances of observing a positron signal from a not-so heavy  dark matter particle.

\begin{figure}[t]
\includegraphics[scale=0.3]{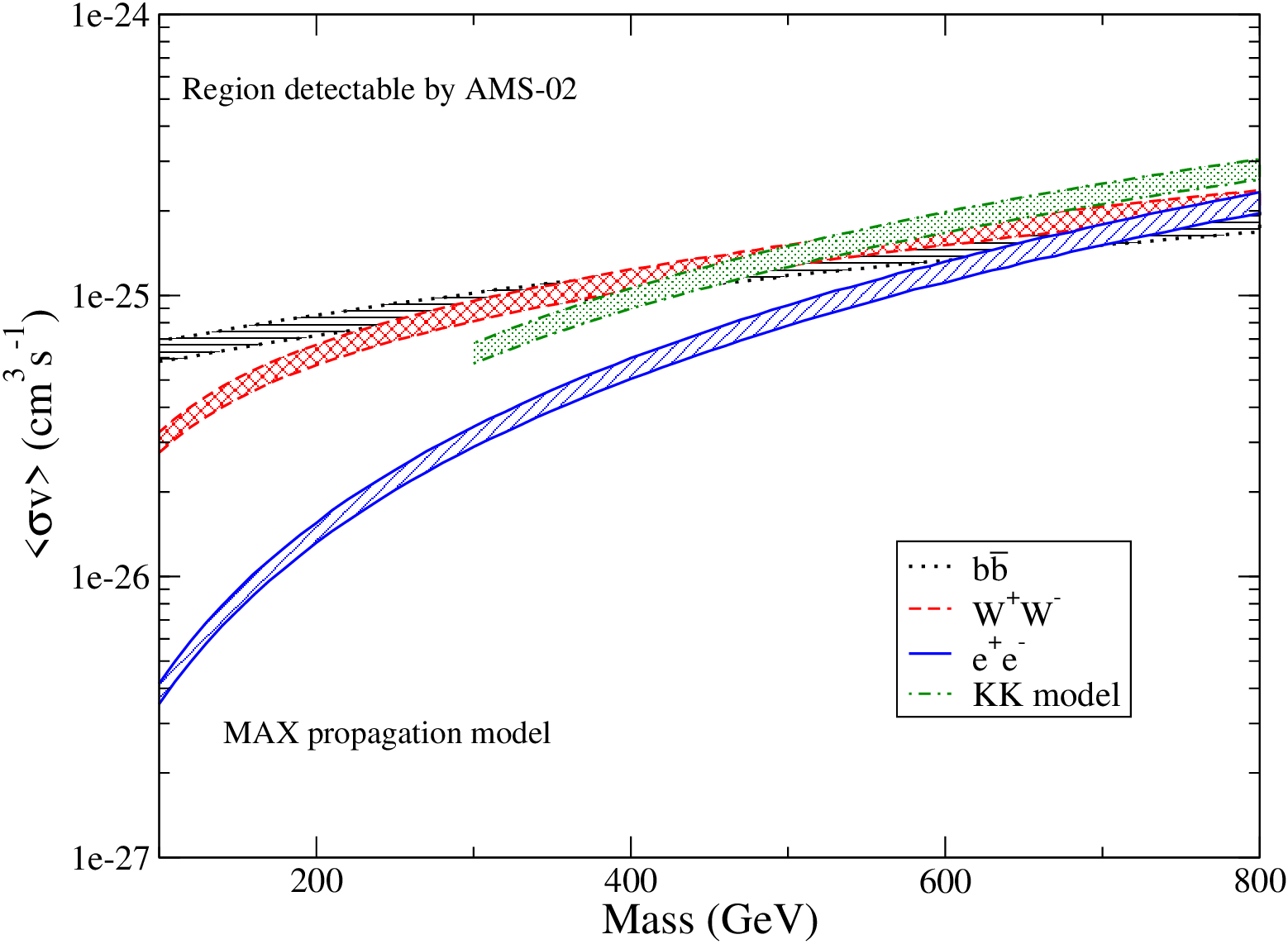} 
\caption{Regions in the plane $(mass, \langle\sigma v\rangle)$ that will be detectable by AMS-02 with three years of data for different dark matter scenarios and the MAX propagation model. The area above the band is detectable for both astrophysical backgrounds. \label{fig:max}}
\end{figure}

Figures \ref{fig:min} and \ref{fig:max} show the detectable regions for the other two propagation models: MIN and MAX. They display the same pattern observed in figure \ref{fig:med} but with slightly larger values of $\sigmav$ for MIN and slightly smaller ones for MAX. These three figures constitute the main result of this paper. They present new detectable regions in AMS-02 for several scenarios of dark matter and different propagation models. 

If a positron excess over the astrophysical backgrounds were indeed observed at AMS-02, it would certainly be difficult to attribute it to dark matter rather than to a more complicated background. To do so, one would likely have to observe a dark matter specific feature such as the sharp spectral edge expected for the model with annihilation into $e^+e^-$, or one would have to combine inputs from collider searches, direct detection experiments, and other indirect detection channels. What we have found in figures \ref{fig:med}, \ref{fig:min} and \ref{fig:max} are the regions in which the positron channel can contribute to the identification of dark matter through indirect searches.

It is important to notice the differences between the  models that are detectable in AMS-02, see previous figures, and those that can  explain the Pamela and Fermi-LAT data in terms of dark matter annihilations. These models require $\sigmav\gtrsim 10^{-23}$cm$^{-3}$s$^{-1}$, multi-TeV masses, and leptonic final states \cite{Meade:2009iu}, and are highly constrained by present data. The detectable models, in constrast, may feature masses in the $100$ GeV range, can annihilate into leptons, quarks, or gauge bosons, and have typical annihilation rates of order $10^{-25}-10^{-26}$cm$^3$s$^{-1}$, small enough to avoid present constraints from antiprotons and gamma rays. Besides, the detectable models include candidates from well-motivated scenarios for physics beyond the standard model, such as universal extra-dimensions and low-energy supersymmetry.

\begin{figure}[t]
\includegraphics[scale=0.3]{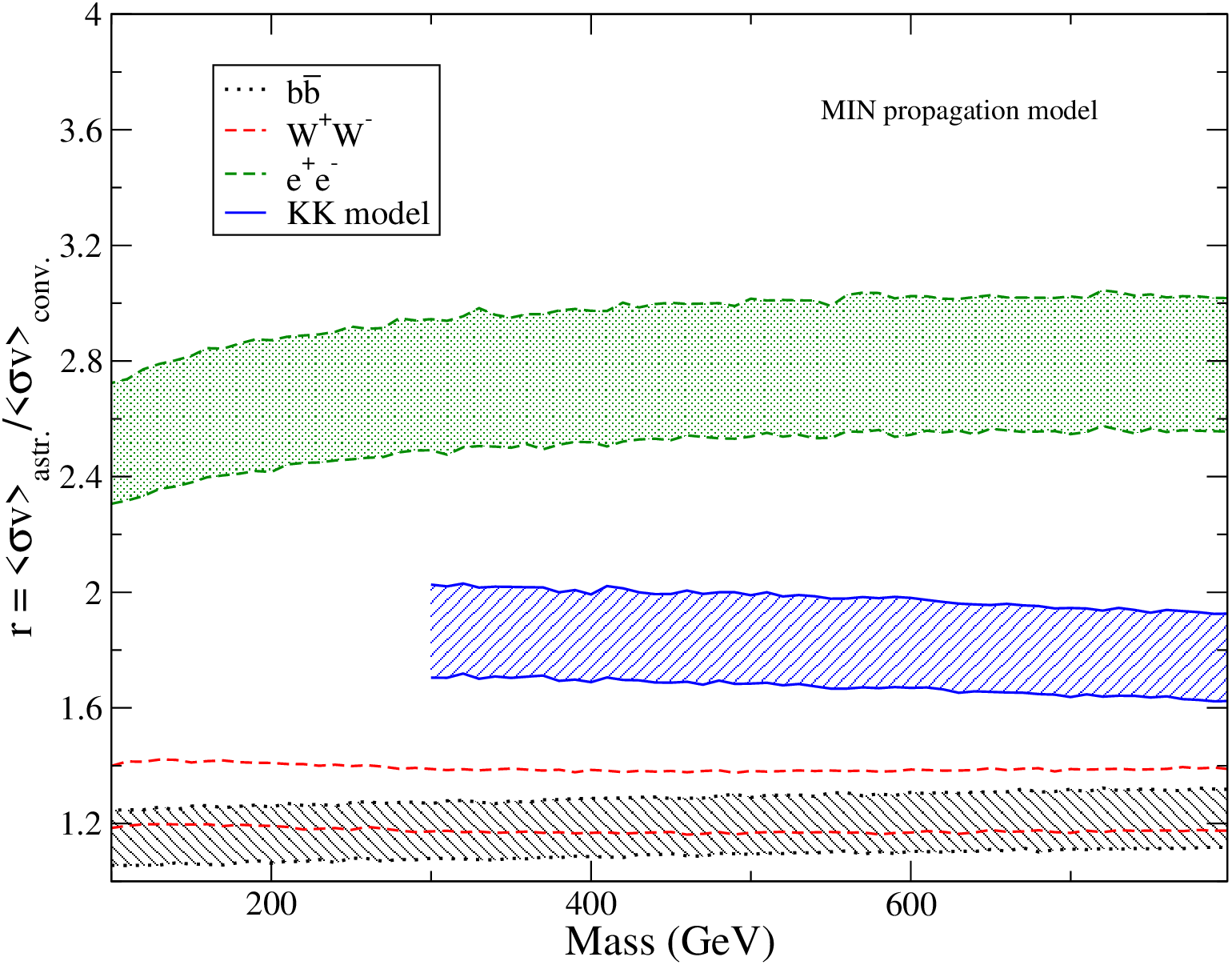} 
\caption{The ratio, $r$, between the annihilation rate required to observe a signal given the astrophysical background and the corresponding rate for the conventional background model.\label{fig:enhmin}}
\end{figure}

One may wonder how much different are the new detectable regions in figures \ref{fig:med}, \ref{fig:min} and \ref{fig:max} from the \emph{old} ones, those obtained for the conventional background model.  Since the astrophysical positron background is larger than the conventional one, we certainly expect that for a given dark matter mass the new regions lie at a higher value of $\sigmav$. How much higher? Let us define a funcion $r$, that tells us exactly that, as
\be
r\equiv\frac{\sigmav_{astr}}{\sigmav_{conv}}\,.
\ee
Here $\sigmav_{astr}$ is the annihilation rate required for detection at AMS-02 given the astrophysical background (the quantity shown in the three previous figures) while $\sigmav_{conv}$ is the corresponding quantity for the conventional background. We know for sure that $r>1$. 
Figure \ref{fig:enhmin} shows $r$ as a function of the dark matter mass for the four dark matter scenarios we consider and the MIN propagation model. As before, the bands correspond to the results for the two astrophysical backgrounds. Notice that the largest value of $r$ we find in the figure is about $3$, for the model with annihilation into $e^+e^-$. Hence, the annihilation rate required to observe a signal is up to  a factor of three larger than the one obtained for the conventional background model. For the KK model $r$ is between $1.6$ and $2.0$, whereas for models with annihilation into $b\bar b$ and $W^+W^-$ is smaller --up to $1.4$. The behaviour of $r$ displayed in the figure can be qualitatively understood from the spectrum (figure \ref{fig:spec}) and the backgrounds (figure \ref{fig:back}). Models with a larger positron flux at higher energies are more sensitive to the difference between the conventional and the astrophysical backgrounds (which is more prominent at higher energies), yielding a larger value of $r$. In other words, a larger $r$ is correlated with a larger positron flux at high energies.

\begin{figure}[t]
\includegraphics[scale=0.3]{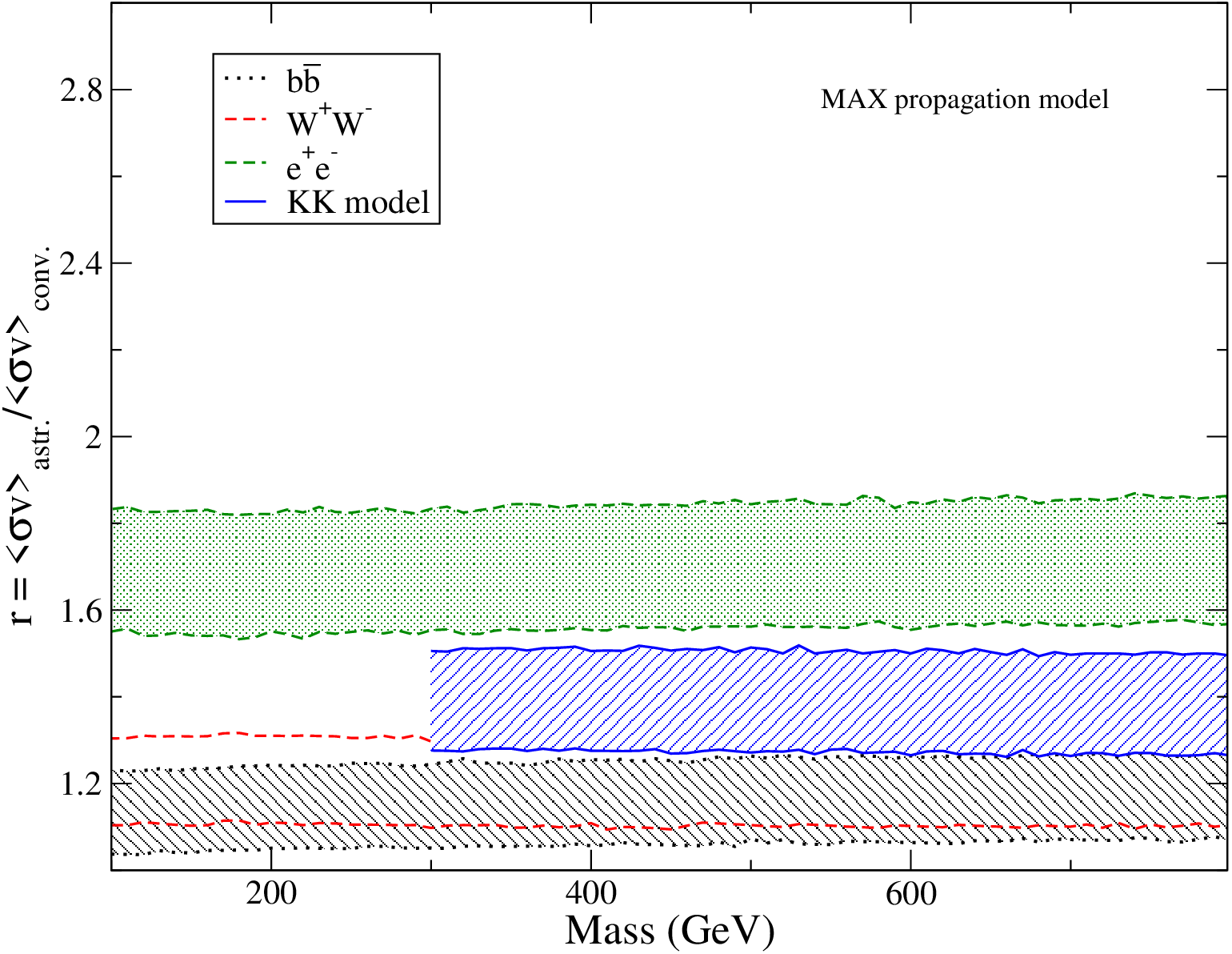} 
\caption{The ratio, $r$, between the annihilation rate required to observe a signal given the astrophysical background and the corresponding rate for the conventional background model.\label{fig:enhmax}}
\end{figure}

In figure \ref{fig:enhmax} we show  $r$ as a function of the dark matter mass for the MAX propagation model. The pattern is the same observed in the previous figure but with smaller values of $r$ for the four models of dark matter. The maximum values of $r$ are approximately given by $1.8$, $1.5$, $1.3$, and $1.3$ respectively for the $e^+e^-$, $KK$, $W^+W^-$, and $b\bar b$ models. As expected, the results for the MED propagation model (not shown) are in between the MIN and MAX results.

These two figures demonstrate the relevant effect of the new astrophysical background on the prospects of detection of a dark matter positron signal in future experiments. Given the  larger background, the signal required to claim an observation is also larger, by up to a factor $3$ for the models we analized. The precise number depends on the dark matter scenario, the propagation model, and the astrophysical background, as illustrated in figures \ref{fig:enhmin} and \ref{fig:enhmax}. 
\section{Conclusion}
We have studied the implications of an astrophysical explanation of the PAMELA and Fermi-LAT data for future searches of positrons from dark matter annihilations. If astrophysical positrons account for such data, they constitute a new background against the positrons from dark matter annihilations. After deriving the new positron background implied by PAMELA and Fermi-LAT, we reassessed the sensitivity of AMS-02 to a positron dark matter signal for several dark matter scenarios and different propagation models. We found the  new detectable regions for AMS-02 and showed that they feature annihilation rates of order $10^{-25}-10^{-26}$ cm$^3$s$^{-1}$.  Hence, despite the larger positron background revealed by PAMELA and Fermi-LAT, the detection of a positron signal from dark matter annihilations is still within the reach of future experiments such as AMS-02.  

\begin{acknowledgments}
We are grateful to David Cerde$\tilde{\textrm{n}}$o  for helpful discussions and encouragement.
C.E. Yaguna is supported by the \emph{Juan de la Cierva} program of the Ministerio de Educacion y Ciencia of Spain. We acknowledge support from  proyecto Nacional FPA2006-01105 and FPA2006-05423, and from the Comunidad de Madrid under Proyecto HEPHACOS, Ayudas de I+D S-0505/ESP-0346. We also thank the ENTApP Network of the ILIAS project RII3-CT-2004-506222 and the Universet Network MRTN-CT-2006-035863. K.Y. Choi is partly supported by the Korea Research Foundation Grant funded by the Korean Government (KRF-2008-341-C00008) and by the second stage of Brain Korea 21 Project in 2006.
\end{acknowledgments}

\end{document}